\documentclass[showpacs,twocolumn,prb,aps]{revtex4}
\usepackage{epsfig}
\usepackage{amsmath}
\usepackage{amssymb}
\begin{document}
\title{Quantum Monte Carlo simulation for the  spin-drag conductance of
the Hubbard model}

\author{Kim Louis and C. Gros} 

\affiliation{Fakult\"at 7, Theoretische Physik,
 University of the Saarland,
66041 Saarbr\"ucken, Germany.}

\date{\today}

\begin{abstract}
In the situation of two electro-statically coupled conductors
a current in one conductor may induce a current in the other one.
We will study this phenomenon, called Coulomb drag,
in the Hubbard chain where the two ``conductors'' are given
by fermions with different spin orientation. 
With the aid of a Monte Carlo (MC) approach
which we presented in a recent paper
we calculate
 the Transconductance in different variants 
of the Hubbard chain 
(with/without impurity and additional [long-ranged] interactions)
 for different fillings.
\end{abstract}
\maketitle

\section{Introduction}
The Coulomb-drag effect describes 
how two conductors (only coupled by the Coulomb force)
 may influence each other.  
Since the Coulomb repulsion is relatively small, a sizeable effect
will only arise when the two conductors are very close to each other.
This condition can be met in  
mesoscopic systems---where with
the advent of new technologies (e.g., carbon nanotubes)
the problem of Coulomb drag
attracted  more and more attention (e.g.,  Refs.\
\onlinecite{Flens,KomnikEgger,NazarAver,Mortensen})---or in the
 spin-drag problem.
For instead of considering  two conductors one may 
look at drag effects between different fermion species, e.g., fermions
with different spin orientation.
Since fermions with different spins are not spatially separated, there
is a large Coulomb force between them which can lead to all kinds of
correlation effects, e.g.,  a drag effect.
In the last years, the interest in  spin-dependent transport increased.
 One key problem is the generation of a
spin-polarized current, i.e., a current where only fermions with one
of the two spin orientations flow. In this context
it is important to keep in mind that the spin-polarized current
may affect the fermions with the opposite spin orientation.
Hence, 
 the drag effect may play here a crucial r\^ole  even though it is
experimentally not directly
accessible. (This is because the driving potentials are in general 
not spin dependent.) 

For this ``spin-drag'' problem  the trans-resistivity of
 higher dimensional systems has been investigated in previous publications, e.g., using
the Boltzmann equation\cite{spindrag1} or the random-phase
approximation.\cite{spindrag2} 
In this paper we focus on the
Transconductance for the Hubbard model.
 While  most authors used
a bosonization approach to compute
the conductance\cite{KaneFish,GogNerTsve}
 we will use
here, for the study of the Transconductance, a Monte Carlo method 
 which we introduced in a 
recent paper.\cite{techpap}
The strategy we followed there was to map 
our fermion  system via the Jordan-Wigner transformation to a spin system which can be analyzed by efficient
though standard Monte Carlo techniques.
We will now extend this method to the one-dimensional Hubbard
model concentrating on the question how a spin-polarized
current (driven by a voltage drop which is assumed to be spin dependent)
 affects the fermions with opposite spin orientation.

In the Section \ref{sec_def} we present the model and give
some central definitions for the spin-drag problem.
The  Sec.\ \ref{MoCar} contains the technical details
on the subject of the MC simulations.
The MC method of our choice\cite{techpap,tpapextract} was 
a variant of the
 Stochastic Series Expansion
 (SSE) as introduced in 
Refs.\ \onlinecite{Sandvik,Sandvik2,Dorneich,SandSyl}.
This method allows an investigation of the one-dimensional Hubbard model.
\cite{SengSand,Clay}

In the following Sec.\ \ref{sec_num_Hubb} we present
our results for the standard Hubbard model and compare with analytical
predictions from bosonization theory.
To obtain a spin-polarized current we
 add in Sec.\ \ref{sec_imp} an impurity to the system, which
acts like a combination of a one-site chemical potential and 
a one-site magnetic
field. We show that such a ``magnetic'' impurity can produce the desired
 spin-polarized current.

The Hubbard model can also be mapped to a 
spinless-fermion ladder.
Hence, our results may also describe that situation,
but there one might argue that the very specific
 modeling of the Coulomb (on-site) interaction in the Hubbard model
is unrealistic (and may differ from other approaches, e.g., Refs.\
\onlinecite{Flens,KomnikEgger,NazarAver,Mortensen}).
We therefore discuss two variants (with additional 
 interaction terms)
 of our model.
First, we will discuss   
in Sec.\ \ref{sec_zig_zag} a situation, 
where fermions with different spins
 live on different sites.
The full system has  the geometry of 
 a zig-zag chain.
%As the Coulomb repulsion in the Hubbard model is relatively strong
%$our calculations differ from the bosonization approaches for mesescopic
%systems 
%(which consider only the effect of the interaction to the low-energy spectrum)
Second, we show in the appendix that a spin-polarized interaction leads
to equal Cis- and Transconductance. This is
 similar to the ``absolute'' drag result
found, e.g., in Ref.\ \onlinecite{NazarAver}.

%In principle the Hubbard model can also describe two coupled 
%conductors. In this picture one assumes that fermions with different spins
%do not live on the same sites. But here the geometrical array of
%the fermion sites can play a crucial r\^ole as is shown
%in Sec.\ \ref{sec_zig_zag}

%%%%%%%%%%%%%%%%%%%%%%%%%%%%%%%%%%%%%%%%%%%%%%%%%%%%%%%%%%%%%%%%%%%%%
\section{Definition of the spin drag in the Hubbard model}
\label{sec_def}

Our model Hamiltonian is the standard Hubbard model (with $N$ sites or atoms)
\begin{eqnarray} H_{\rm Hubb}&=&t\sum_n\sum_{\sigma=\uparrow,\downarrow}
\left(c_{n,\sigma}^\dag c_{n+1,\sigma}+c_{n+1,\sigma}^\dag c_{n,\sigma}\right)\nonumber\\
&&-\mu\sum_{n,\sigma} n_{n,\sigma}
+U\sum_n n_{n,\uparrow}n_{n,\downarrow},\label{Hhubb}\end{eqnarray}
where $n_{n,\sigma}$ is the occupation number of fermions with spin
$\sigma$ at site $n$, and $c_{n,\sigma}^{(\dag)}$ is the corresponding
annihilation (creation) operator.
To perform our  transport calculations, 
we will use  the approach from Ref.\  \onlinecite{techpap}.
  Since the hopping term does not connect fermions with different spins,
 it is natural
to consider current and potential operators for each spin orientation 
separately. Following Ref.\  \onlinecite{techpap}
 the potential operators read then: ($e$ being the charge unit and $x$
being  the position of the voltage drop)
$$P^{\downarrow}_x=e\sum_{n>x}n_{n,\downarrow},\qquad
P^{\uparrow}_x=e\sum_{n>x}n_{n,\uparrow}. $$
The conductance (of a spinless-fermion chain)
is the linear response of one potential operator
to another; therefore the explicit form of the
current operators is not needed here.\cite{techpap}
As we have two potentials, we can define four transport quantities
(conductances)
$g_{ij}$ which describe the (linear) response of $P^i$ to $P^j$ where $i,j\in
\{\uparrow,\downarrow\}$. 
Further details on how to evaluate the
$g_{ij}$'s
are to be found in section \ref{MoCar}.

For the moment we will discuss only symmetric models, i.e., we have
spin-rotational invariance (the only asymmetric  model
that we will discuss appears in Sec.\ \ref{sec_imp}); thus, we  end up
with only two distinct quantities.
We call $g_c:=g_{{\downarrow}{\downarrow}}=g_{{\uparrow}{\uparrow}}$ the {\em Cisconductance}
%as the response of $P^{\uparrow}_x$ to $P^{\uparrow}_y$ 
%(or likewise the response of $P^{\downarrow}_x$ to $P^{\downarrow}_y$),
 and 
$g_t:=g_{{\downarrow}{\uparrow}}=g_{{\uparrow}{\downarrow}}$
the 
{\em Transconductance}.

The naming conventions come from  
the physical interpretation of these coefficients which is the following:
If we  switch on at a certain time a (supposedly) spin-polarized
potential which acts only on spin-up fermions
(i.e., we add a time-dependent perturbation 
 of the form $$VP^\uparrow\theta(t)$$ with $V$ being the
voltage amplitude and $\theta$ the Heavyside-step function)
 we will find a current of
 spin-up fermions %(also referred to as ``up-spins'')
$$I^\uparrow=g_{c}V.$$
This is the {\em drive} current governed by the Cisconductance,
but there will also be a current of spin-down fermions %(also referred to as  ``down-spins'') 
$$I^\downarrow=g_{t}V,$$
the {\em drag} current (governed by the Transconductance).
The latter may be nonzero,
even though the spin-down fermions do not feel the applied potential.

%as the response of $P^{\cal E}_x$ to $P^{\cal O}_y$ 
%(or likewise the response of $P^{\cal O}_x$ to $P^{\cal E}_y$).
The situation of a nonvanishing Transconductance (or drag current)
is called {\em Coulomb drag}. This problem has been studied, e.g.,
for coupled spinless-fermion systems
by bosonization in
Refs.\ \onlinecite{Flens,KomnikEgger,NazarAver} and to second-order
 perturbation theory in Ref.\ \onlinecite{Mortensen}.

 Normally, (since  spin-polarized potentials are not available)
 one is only interested in the (full) conductance
of the Hubbard model, where both 
fermion species  feel the same potential,
 and the full current is the sum of
the currents of the
spin-up and spin-down fermions.
As is straightforward to see, the Cis- and Transconductance give us directly
the conductance of the Hubbard model
via the relation $g_{\rm Hubbard}=2(g_c+g_t)$.
%%%%%%%%%%%%%%%%%%%%%%%%%%%%%%%%%%%%%%%%%%%%%%%%%%%%%%%%%%%%%%%%%%%%%%%%%%
\section{Description of the Monte Carlo method}
\label{MoCar}
%\subsection{SSE}
Before starting with Monte Carlo we have
to cast our Hamiltonian in a convenient form.

\begin{figure}
\epsfig{file=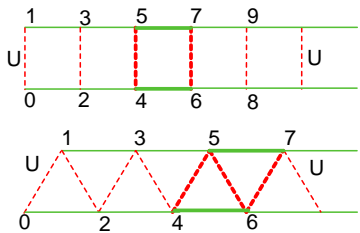,width=0.26\textwidth,angle=0}
\caption{ The model Hamiltonian that will be discussed in this
 paper. Solid lines indicate a full Heisenberg-like
 interaction between the sites; dashed lines stand for sites coupled 
only by a $z$-$z$ term (Ising-like interaction). 
The upper part is for $U^\prime=0$.
The lower for $U^\prime=U$. (Other values of $U^\prime$ are not
 considered.) A {\it plaquette} as used in the MC scheme is indicated by
boldface lines.
}
\label{hsplit}\end{figure}

Using the Jordan-Wigner transform the Hubbard model can be mapped
to a spin ladder.
To each occupation operator we introduce a spin operator, i.e.,
we replace
 $n_{n,\uparrow}\to S_{2n}^z+1/2$ and
$n_{n,\downarrow}\to S_{2n+1}^z+1/2$. If we express the Hubbard Hamiltonian
with those new operators we obtain the following spin ladder (with $2N$ sites):

%The system that we will study is quasi one-dimensional:
%It 
% consists of two $z$-$z$-coupled $xxz$ chains (with altogether $N$
% sites)
%and is essentially the
% Hubbard model, if one interprets the two chains as fermions with
% different spin indices 
%The Hamiltonian reads explicitly: 
\begin{eqnarray}
H\!&=&\!\!\sum_{n}
 \left[J_x(S_{n}^+S_{n+2}^-+S_n^-S_{n+2}^+)/2+J_zS_n^zS_{n+1}^z+BS_n^z\right]
\nonumber \\
&+&\sum_n \left[US_{2n}^zS_{2n+1}^z+U^\prime S_{2n+1}^zS_{2n+2}^z\right],
\label{hamillad}
\end{eqnarray}
where the  sites with even number represent spin-up fermions
and the  sites with odd numbers, spin-down fermions. 
(For the Hubbard model one has to put
$J_z=0=U^\prime$. These parameters are used to model the spin-polarized
interaction from the appendix and the
zig-zag chain from Sec.\ \ref{sec_zig_zag};
see bottom half of Fig.\ \ref{hsplit}.
The sites $2n$ and $2n+1$ in the ladder represent therefore one atom of the
Hubbard model and interact via an Ising interaction representing
the Coulomb force
(see  Fig.\ \ref{hsplit}). The hopping amplitudes satisfy
$J_x=2t$, and the strength of the magnetic field $B$ in 
Eq.\ (\ref{hamillad}) is obtained from the chemical potential
via the relation $B=U/2+U^\prime/2-\mu$.
% if expressed with the parameters of the Hubbard model.
Half filling corresponds therefore to $B=0$.
The two potential operators $P_x^{\uparrow,\downarrow}$
 introduced in the previous section
can be related to
potential operators for the two chains ($e$ being the charge unit)
$$P^{\downarrow}_x=e\sum_{n>x}S^z_{2n+1},\qquad P^{\uparrow}_x=e\sum_{n>x}S^z_{2n}. $$
%The conductance (of one chain)
%is the linear response of one potential operator
%to another; \cite{techpap} but
%in the case of two coupled chains 
Then we may obtain the four conductances $g_{ij}$
introduced in the previous
 section
 by computing
($i,j\in\{\downarrow,{\uparrow}\}$)
\begin{equation}\label{intbypart2}g_{ij}(\omega_M)=-\omega_M/\hbar\int_0^{\hbar\beta} \cos(\omega_M
\tau)\langle P_x^iP_y^j(i\tau)\rangle d\tau
\end{equation}
at the Matsubara frequencies $\omega_M=2\pi
M(\beta\hbar)^{-1},\; M\in{\mathbb N}$, and 
then extrapolating to $\omega=0$.
(The extrapolated value should not depend on $x$ or $y$.\cite{techpap}
We chose $x=N/2=y-1$.)
For the extrapolation from $g(\omega_M)$ to $g(\omega=0)$ we will use a
quadratic fit from the first three Matsubara frequencies.
[We will use open (OBC's) instead of periodic boundary conditions
(PBC's).]
Since the Hamiltonian contains Heisenberg-like interactions as well as
Ising-like interactions, %one will run into problems (namely, an enhanced bounce weight) %when implementing an
%SSE-version for this Hamiltonian.
%In this respect the
% situation is similar to the frustrated Hamiltonian discussed
%in Ref.\ \onlinecite{tpapextract}.
%Hence 
it is advantageous to use the Stochastic Cluster Series Expansion
(SCSE) introduced in Ref.\
\onlinecite{tpapextract}. For the Hubbard model
the SCSE gives essentially the same update scheme as the one used in Ref.\
\onlinecite{SengSand}.
We will explain it now shortly.

Following Ref.\ \onlinecite{tpapextract} we
 split the Hamiltonian according to $H=\sum_{h\in \frak{h}}h$,
but this time  %the set $\frak h$ consists
into four-sites clusters, called plaquettes  (see  Fig.\ \ref{hsplit}).
The following operators belong to the plaquette $\cal P$
(containing interactions between the sites $2n$, $2n+1$, $2n+2$, and $2n+3$):
\begin{eqnarray*}
h_{\cal P}^{(1)}&=&J_xS_{2n}^+S_{2n+2}^-/2\qquad\quad
h_{\cal P}^{(2)}=J_xS_{2n}^-S_{2n+2}^+/2\\
h_{\cal P}^{(3)}&=&J_xS_{2n+1}^+S_{2n+3}^-/2\qquad
h_{\cal P}^{(4)}=J_xS_{2n+1}^-S_{2n+3}^+/2\\
h_{\cal P}^{(5)}&=&C+U/2S_{2n}^zS_{2n+1}^z
+U/2S_{2n+2}^zS_{2n+3}^z+\\
&+&U^\prime S_{2n+1}^zS_{2n+2}^z+J_{z}S_{2n+1}^zS_{2n+3}^z+
J_{z}S_{2n}^zS_{2n+2}^z\\
&+&B/2(S_{2n}^z+S_{2n+1}^z+S_{2n+2}^z+S_{2n+3}^z). \\
\end{eqnarray*}
The set $\frak h$ consists of all $h_{\cal P}^{(t)}$ for all plaquettes
$\cal P$ and all $t=1,\dots, 5$.

The heart of the SCSE program is the so called loop update, where
 a spin flip of a subset (loop)
of all spin variables is proposed.
Since the sites with even and  odd numbers form two chains, 
which are only coupled by a $z$-$z$ interaction term, 
we find that the set of spin variables that will be flipped 
in the loop update
belongs entirely to one of the chains.
Therefore, we can view the new algorithm as
 {\em making loop updates for each chain separately}.
During a loop update for one chain  the spin variables of the other chain
 remain fixed. The consequence of this is that,
 if we update, e.g., the even chain, then operators with superscript
 $i=3,4$ can be  neglected (are irrelevant for the loop construction),
%the term $J_{z}S_{2n+1}^zS_{2n+3}^z$ in the operators with superscript
% $i=5$ gives an irrelevant constant,
and the coupling terms (between the chains) reduce to  magnetic field
terms (for the even chain).

It is however advantageous to consider another variant of the
loop update. The construction is similar to the first variant, but now we propose
spin flips for both chains, i.e., the spin variables belonging to
sites $2n$ and $2n+1$ are flipped simultaneously.
 This may be viewed as
a construction of  two parallel loops---one for each chain.
 Since the two loops must be parallel, the number of
possible transitions between different plaquette states is reduced. 
This may lead to a less efficient algorithm,\cite{SandSyl}
but one should note that this parallel-loop 
update becomes deterministic for the case
of $B=0$ and hence enhances the efficiency of the algorithm 
(at least for this situation) considerably.

%%%%%%%%%%%%%%%%%%%%%%%%%%%%%%%%%%%%%%%%%%%%%%%%%%%%%%
\section{Numerical results}
\label{sec_num_res}

\vspace{-0.5cm}%%%%%%%%%%%%%%%%%%%%%%%
\subsection{Transconductance in the Hubbard model}
\label{sec_num_Hubb}
\subsubsection{ Comparison with the Hubbard model}
\label{sec_hubb}
%We first consider the model with $U^\prime=0$.
%For this case our Hamiltonian maps via the 
%Jordan-Wigner transformation
%to the one-dimensional Hubbard model.
%This transformation maps $S_{2n}^z\to n_{n,\uparrow}-1/2$ and 
%$S_{2n+1}^z\to n_{n,\downarrow}-1/2$ where $n_{n,\sigma}$
%is the occupation number of fermions on  site $n$
%[of the fermionic chain] with
%spin index $\sigma$.

%From the perspective of the
% Hubbard model, the two chains (${\uparrow}/{\downarrow}$) represent fermions with different spin
% indices (up/down).
%Computing the Cis- and Transconductance may therefore also be interesting for spintronics.

% We found $g_{\rm Hubbard}=0$ from $g_c+g_t\to 0$
%which is unfortunately in sharp contrast with the bosonization result
%$g_{\rm Hubbard}=K_\rho \cdot 2e^2/h$\cite{Ogata} where the Luttinger
%parameter assumes the value $K_\rho=1/2$ at 
%half filling {\em for any positive $U$}.\cite{Schulz}
%This discrepancy is possibly a finite temperature effect.

In bosonization theory the Hubbard model is described by two boson fields
$\Phi_{\uparrow,\downarrow}$ representing the degrees of freedom of 
different spin orientations.
The current operators for the spin sectors are then given by $J_{\uparrow,\downarrow}\propto
\partial_x \Phi_{\uparrow,\downarrow}$.\cite{KaneFish}
The conductance can be written in the form of
 a current-current correlator\cite{KaneFish}
and may be evaluated 
in terms of the Luttinger-liquid parameters $K_{\rho,\sigma}$
of the charge and spin field\cite{GogNerTsve}
$\Phi_{\rho/\sigma}=(\Phi_\uparrow\pm\Phi_\downarrow)/\sqrt{2}$.
% It is advantageous to introduce 
%the charge and spin field
%\begin{equation}\Phi_\rho=\frac{1}{\sqrt{2}}(\Phi_\uparrow+\Phi_\downarrow)
%\qquad \Phi_\sigma=\frac{1}{\sqrt{2}}(\Phi_\uparrow-\Phi_\downarrow),
%\label{Phirs}\end{equation}
%as the system may then be described by two Luttinger liquids with
%Luttinger parameters  $K_{\rho,\sigma}$.\cite{GogNerTsve}
%Since the conductance is given by a current-current correlator\cite{KaneFish}
%Eq. (\ref{Phirs}) translates to a relation between the Luttinger
%parameters and the two conductances, namely, 
The result is (using the linearity of the correlator and the results
from Ref.\ \onlinecite{ApelRice})
\begin{equation}g_c=\frac{1}{2}(K_\rho+K_\sigma)\qquad g_t=\frac{1}{2}(K_\rho-K_\sigma).\label{Krs}\end{equation}

When $\mu=U/2$ we are at half filling, where
 Umklapp processes are responsible for a gap in the system.\cite{Schulz}
The (charge) gap $\Delta(U)$ depends on the Hubbard repulsion $U$
and is finite for all $U$.

\subsubsection{Numerical simulations}
We present now Monte Carlo results for
 the Transconductance in the Hubbard model Eq.\
(\ref{Hhubb}) [or Eq.\ (\ref{hamillad}) for $U^\prime=0$].% and compare our results with the bosonization results.

%The reader should recall that
%applying Jordan and Wigner's transformation the inter-chain coupling in the
%spin system Eq.\ (\ref{hamillad}) maps to a dressed interaction in the fermionic system, i.e.,
% a Coulomb repulsion with an additional
% shift in chemical potential that will save the system from depletion.

We performed simulations   for  two different 
chemical potentials: First, $\mu=U/2$ corresponding to half filling
and second $\mu=0$.
% corresponding to the chemical potential shift induced by the Coulomb interaction.
%representing a simple inter-chain (Coulomb) interaction without any chemical potential shift.
In the latter case
 the system is no longer at half filling, but has a $U$-dependent
  filling, which is shown in
Fig.\ \ref{stgmU} of the appendix (in the large-$U$ limit the system
reaches quarter filling). %In the large-$U$ limit the 
%Here we should like to note that the bosonization results
%cited above are done for the finite $B$-case.

\begin{figure}
\epsfig{file=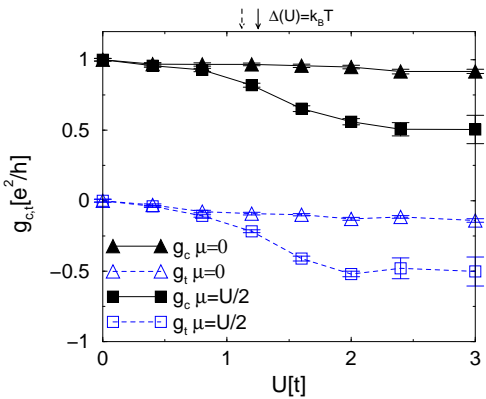,width=0.35\textwidth,angle=0}
\caption{Cis- and Transconductance (filled/empty symbols)
of the Hubbard model as a function of $U$ for two different $\mu$'s 
(200 sites, $T=0.02t/k_B$, OBC's, $2\cdot
 10^5$ MC sweeps.) The (solid) arrow indicates the $U_T$ for which 
 the charge gap
(present at half filling) satisfies
$\Delta(U_T)=k_BT$. (The dotted arrow shows $U_{T/2}$ where $\Delta(U_{T/2})=k_BT/2$ for comparison.) }
\label{hullz1a}\end{figure}

We show $g_c$ and $g_t$ as a function of $U$
for the two different $\mu$'s 
in Fig.\ \ref{hullz1a}. 
%and  Fig.\ \ref{hullB1z1} ($\mu=0$).
 The figure shows that the Coulomb drag is very sensitive to a
change in chemical potential.

Let us first look at the {\em half-filled} case (Fig.\ \ref{hullz1a}). If $U$
is very large the Coulomb repulsion acts as an effective projection
to those configurations satisfying  $P^{\downarrow}=-P^{\uparrow}$.
This implies that
 $g_c+g_t\to 0$ as $U\to \infty$. This contemplation is
in accordance with Fig.\ \ref{hullz1a}.
We should actually  expect from Eq.\ (\ref{Krs})
that for $T=0$ we have $g_c+g_t=K_\rho=0$ 
 because of the charge gap $\Delta(U)>0$ for all $U>0$
(cf.\ Ref.\ \onlinecite{Schulz}).
This should lead to a discontinuous jump at
 $U=0$, because without the Coulomb force 
  evidently  $g_t=0$ and $g_c$ is the 
 conductance of
uncoupled spinless-fermion chains from Ref.\ \onlinecite{ApelRice}.
Here we emphasize
 that our method is a finite-temperature method, 
which means that the conductances calculated by us interpolate
smoothly between the values for $U=0$ and  $U=\infty$.
The crossover  is expected to take place at that interaction value $U_T$
which satisfies $\Delta(U_T)=k_BT$.
It is therefore interesting to see how $g_{c,t}$ scale with temperature.
However,
our method gives only access to the low-$T$ regime,\cite{techpap} 
such that we will compare here results for only two different temperatures,
$k_B T=0.01t$ and $k_BT= 0.02t$.  
In the two subsequent simulations for the Cis-/Transconductance
we did not find any difference at all.
 This implies only a  weak temperature dependence
(at low $T$) for the  interaction $U_T$ which governs
the crossover. Since $\Delta(U)$ is known from analytical 
results,\cite{klumper}
we can calculate  the two crossover
interactions---defined by $\Delta(U_{0.02})=0.02 t$ and
$\Delta(U_{0.01})=0.01 t$---finding that
 $U_{0.02}\approx 1.25t$ and $U_{0.01}\approx 1.12t$ do not differ much,
 as expected (they are also both
indicated by arrows in Fig.\
\ref{hullz1a}).

%This in accordance with the fact that our (finite temperature) prediction
%for an effective 
%vanishes in Fig.\ \ref{hullz1a} in the limit $U\to\infty$
%even for  the $J_{z}=0$-case.
Another important consequence of $g_c+g_t\to 0$
 is that the signs of the Cis- and 
Transconductance are opposite or---in terms of the spin-up and spin-down
currents---that the induced drag current
flows in the opposite direction of the drive current.\cite{Schlott}

Now we will turn to the spin sector.
We have 
 $K_\sigma=1$ by spin-rotational invariance of the Hubbard model\cite{Schulz}
 implying [see Eq.\ (\ref{Krs})] $g_c-g_t=1$ for all $U$  which is very well satisfied
by Fig.\ \ref{hullz1a}.

Putting the two results for $g_c\pm g_t$ together, we obtain 
$g_c=0.5\, e^2/h=-g_t$ valid at high $U$. This large-$U$ limit
of $g_c$ may also be computed in
second-order perturbation theory. In this approximation 
 the Hubbard model can be mapped to a Heisenberg model.
The operator [on the Hilbert space of the original Hamiltonian Eq.\
 (\ref{Hhubb})] $P^{\uparrow}_x$, which is effectively equal to $-P^{\downarrow}_x$,
is identified 
with the operator (on the Hilbert space of the effective Heisenberg
 model)
$P_{x}=\sum_{n>x}T_n^z$.
 [ $T_n^z=(n_{n,\uparrow}-n_{n,\downarrow})/2$ is the spin operator for
 fermions; here denoted by $T$ to avoid confusion with the spin
 operators appearing in  Sec.\ \ref{MoCar}.] Applying the results from
 Ref.\ \onlinecite{techpap}
 the computation of $g_c$
 reduces then to the computation of the spin conductance of the Heisenberg
 model, which equals one half in units $e^2/h$.

In the case of {\em zero} $\mu$ (again Fig.\ \ref{hullz1a})
where the system is away from half filling there is no charge gap.
Hence, $K_\rho $ is finite, and so 
Eq.\ (\ref{Krs}) tells us that $g_c+g_t=K_\rho$ does not decay with $U$.
[Here we note that 
$g_c+g_t$ agrees (within error bars) with the values for
$K_\rho$ available in Ref.\
\onlinecite{Schulz}.]
We still have  $K_\sigma=1$, which leads 
to 
$g_c-g_t=1$ for all $U$ (again very well satisfied by the figure).

Finally, we  consider the large-$U$ limit. 
%Here, at low $T$ the conductance is $e^2/h$.
Inserting  $K_\rho(U=\infty)=0.5$ and $K_\sigma\equiv 1$
from Ref.\ \onlinecite{Schulz}
%With the relation $g_c-g_t=1$ (from spin-rotational invariance)
 in Eq.\ (\ref{Krs}) yields $g_c=0.75$ and $g_t=-0.25$ (units $e^2/h$).
%which agree
%with the large $U$-limit $K_\rho=0.5$ from Ref.\ \onlinecite{Schulz}.
These results
are in accordance with the figure.
(Note that the statistical error increases with $U$ such that we cannot
compute $g(U)$ for sufficiently high $U$ in order to extract the
large-$U$ limit accurately.)

%%%%%%%%%%%%%%%%%%%%%%%%%%%%%%%%%%%%
\subsection{Magnetic Impurity}
\label{sec_imp}
\begin{figure}
\epsfig{file=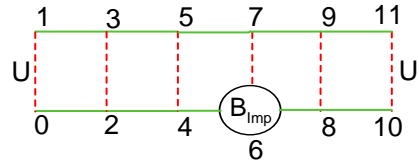,width=0.3\textwidth,angle=0}
\caption{ The model Hamiltonian with an impurity
at site $N/2$ for $N=6$ in the spin ladder representation from 
the upper half of Fig.\ \ref{hsplit} (for $U^\prime=0$). 
The site on which the impurity potential acts is encircled.
}
\label{hsplitb}\end{figure}

In this subsection we will study the influence of
an impurity. The modeling of the impurity follows
Ref.\ \onlinecite{KaneFish}, but for the spin-drag problem it is 
natural to consider a spin-dependent impurity, as we will do here.
We extend therefore our Hamiltonian in the following way 
$$H=H_{\rm Hubb}+B_{\rm Imp}n_{N/2,\uparrow}\;,$$
i.e., we introduce an (impurity) potential at exactly one central site 
 (which acts only on one spin orientation, see Fig.\ \ref{hsplitb}).
\begin{figure}[h]
\epsfig{file=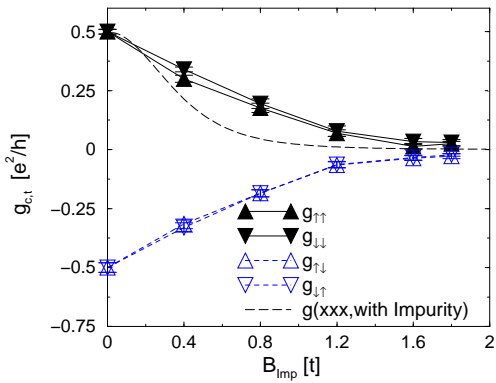,width=0.36\textwidth,angle=0}
\caption{The Cis- and Transconductance (filled/empty symbols) as a function 
of the impurity strength
 $B_{\rm Imp}$ at half filling. The impurity acts on spin-up fermions.
Note that the Cisconductance
 for the spin-up fermions
 $g_{\uparrow\uparrow}$ (triangle up) might differ from the 
 one of the spin-down fermions $g_{\downarrow\downarrow}$ (triangle down).
The two Transconductances are the same.
The conductance of an  $xxx$ (Heisenberg) chain 
 with one  impurity (which should coincide with the
 large-$U$ limit of the Cisconductance) is given for comparison.
($\mu=U/2$, $U=2t$, $N=192$ sites, $T=0.02t/k_B$.)
}
\label{hull1KF}
\end{figure}
\begin{figure}[h]
\epsfig{file=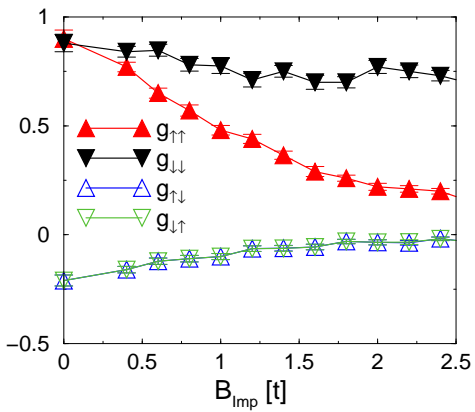,width=0.35\textwidth,angle=0}
\caption{ The Cis- and Transconductance (filled/empty symbols) as a function 
of the impurity strength
 $B_{\rm Imp}$ in the Hubbard model away from half filling. The impurity acts on spin-up fermions.
Note that the Cisconductance
  for the  spin-up fermions
 $g_{\uparrow\uparrow}$ (triangle up) differs from the 
one of the spin-down fermions $g_{\downarrow\downarrow}$ (triangle down).
The two Transconductances are the same.
($\mu=0$, $U=4t$, $N=192$ sites, $T=0.02t/k_B$.)
}
\label{hull2BKF}
\end{figure}

Fig.\ \ref{hull1KF} 
shows Cisconductances and
Transconductances as a function of the impurity potential $B_{\rm
Imp}$ at {\it half filling}.
(The example is chosen such that the Transconductance
in the unperturbed system is relatively large.)
The conductance of the Heisenberg chain with one impurity,
which is the large-$U$ limit of $g_c$, is given 
for comparison.

Although the two Cisconductances, $g_{\uparrow\uparrow}$ and
$g_{\downarrow\downarrow}$, could in principle
differ
(the model is now asymmetric)
they do not in the case of half filling---at least within error bars.
Both Cisconductance and Transconductance go---more or less linearly---to zero as the
impurity strength increases.

We note that within error bars $g_c=-g_t$ such that 
the full conductance of the system
$$g=2(g_c+g_t)$$ remains zero after insertion of the impurity.
Furthermore, investigations with our method at different temperatures 
find no sizeable $T$-dependence.

\begin{figure}[h]
\epsfig{file=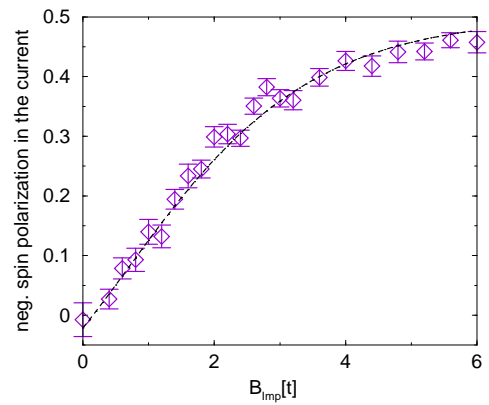,width=0.35\textwidth,angle=0}
\caption{ The negative average spin polarization ($-\cal S$) of the induced
current as a function of the impurity strength
 $B_{\rm Imp}$ in the Hubbard model away from half filling. The impurity
 acts on spin-up fermions. The dashed curve is obtained by 
fitting $g_{\uparrow\uparrow}$, $g_{\downarrow\downarrow}$
and $g_t$ in Fig.\ \ref{hull2BKF} with an exponential ansatz and 
substituting these fits  into the formula for $\cal S$.
($\mu=0$, $U=4t$,  $N=192$ sites, $T=0.02t/k_B$.)
}
\label{spinpol}
\end{figure}

In the case of {\it zero chemical potential,  $\mu=0$,} we find a splitting of the two
Cisconductances  (see Fig.\ \ref{hull2BKF}).
This is particularly interesting since this finding implies a
spin-polarized current. If we assume a (spin-independent) driving potential of the form 
$$V(P^\uparrow+P^\downarrow)$$
($V$ being the voltage amplitude),
then the current is $$I=I^\uparrow+I^\downarrow, \qquad 
I^\uparrow=(g_{\uparrow\uparrow}+g_{\uparrow\downarrow})V
\quad I^\downarrow=(g_{\downarrow\uparrow}+g_{\downarrow\downarrow})V.$$
The average spin of a fermion in the current is therefore
(using $g_{\downarrow\uparrow}=g_{\uparrow\downarrow}=g_t$)
$${\cal S}=\frac{I^\uparrow-I^\downarrow}{2I}=
\frac{g_{\uparrow\uparrow}-g_{\downarrow\downarrow}}{2(g_{\uparrow\uparrow}+g_{\downarrow\downarrow}+2g_t)}$$
different from zero (see Fig.\ \ref{spinpol}).

An interesting problem is the question
whether for $\mu=0$ the Cisconductance in the pure (spin-down) sector
$g_{\downarrow\downarrow}$
survives or not when we increase $B_{\rm Imp}$ to infinity.
Note that a finite $g_{\downarrow\downarrow}$
would imply a total spin polarization, i.e., ${\cal S}=-1/2$.
The limit  $B_{\rm Imp}\to \infty$
 of  $g_{\downarrow\downarrow}$
 cannot be taken directly
(because of problems with the MC simulation), but here we note
that there is another way to model the impurity.
Instead of applying a local magnetic
field on one site, one can 
introduce a weak link, i.e., decrease
the hopping amplitude for spin-up fermions
 between the sites $N/2$ and $N/2+1$ from the
initial value $t$ to $t_{{\rm Imp}}^\uparrow$.
These two variants of impurities behave similarly.\cite{KaneFish}
We computed the Cisconductance for the unaffected spin orientation
 for the model
with $t_{{\rm Imp}}^\uparrow=0$ (corresponds to $B_{\rm Imp}=\infty$)
at $\mu=0$, $U=4t$, $N=200$ sites.
 We find a value of about $[0.79\pm 0.03]{ e^2/h}$.

The different behavior of the Cisconductance in the (unaffected) spin-down
sector at
half filling and away from half filling may be explained as follows:
Suppose $B_{\rm Imp}$ and $U$ are large.
%We use the fermionic picture introduced in Sec.\ \ref{sec_hubb}.
%There 
The effect of the $B_{\rm Imp}$ term on the fermions is that it
 forbids occupation of the impurity site 
for one of the two fermion species (in our case spin-up fermions).
At half filling a spin-up fermion  can hop only from one site to another by 
 exchanging the site with a spin-down fermion
 (there are no empty sites), i.e., 
simultaneously with the spin-up  a spin-down fermion must hop
 in the opposite direction {\it et vice versa} (implying $g_c=-g_t$).
A fermion of a certain spin index can then only pass the impurity site 
if accompanied by
 a fermion of opposite spin (which moves in the opposite direction).
Since the impurity site is forbidden for one of the two fermion species,
no fermion can pass the impurity site, and
both Cisconductances must go to zero. 

Away from half filling the hopping of an spin-up fermion
does not necessarily require the hopping of a spin-down fermion
 (the spin-up fermion can
hop to an empty site) and hence the impurity affects only one of the two
Cisconductances.
%%%%%%%%%%%%%%%%%%%%%%%%%%%%%%%%%
\subsection{Zig-zag chain}
\label{sec_zig_zag}

So far we have dealt with 
a system of two fermion species,  where the two
species reside on the same set of sites.
In contrast to this,
the bosonization approaches considered mostly systems
of two coupled spinless-fermion  conductors.
We can compare our results
with that situation, if we interpret
the Hubbard model as a spinless-fermion ladder.
Here one assumes that
each fermion species lives on a different conductor
[i.e., the two indices ($n,\uparrow$) and ($n,\downarrow$) are supposed to
label (spatially) different sites; compare Sec.\ \ref{MoCar} and 
upper half of Fig.\ \ref{hsplit}].
But one should note that for this case the parameter $U$ should be small
(since the distance between separated conductors is large) 
and the Coulomb interaction should be long-ranged (not on-site as in the
Hubbard model). Hence we are led to the question
how  a variation in the interaction term modifies our results.

%Since the interaction between two sites depends on their distance.
%The geometry of the configuration plays here a major role.
%As a first step  towards a  long-ranged interaction,
%with the geometry of a zig-zag chain,
%where the fermion species live on different sites,
%In principle such a system can also be studied using the Hubbard model,
%if one assumes that the two indices $n,\uparrow$ and $n,\downarrow$
%label (spatially) different sites. But if we do so, we need to justify why
%we  only keep the interaction between sites $n,\uparrow$ and
%$n,\downarrow$
%and not between other sites. 
%Thus one is led to the following question:  Which influence does
% the geometrical array of sites have on the Coulomb drag?
%We will address this question by considering  a model
%where
To address this question we  add a new interaction term to the Hubbard model
% where each spin-down site lies  between two spin-up sites 
%yielding a 
% system with the geometry of a (frustrated) zig-zag chain
% and the Hamiltonian:
$$H=H_{\rm Hubb}+U\sum_nn_{n,\downarrow}n_{n+1,\uparrow}\;.$$
This Hamiltonian corresponds to Eq.\ (\ref{hamillad}) with $U^\prime=U$.
%(see lower half of Fig.\ \ref{hsplit}).
One can justify introducing this new interaction term 
 if the fermion species live on different sites where
each spin-down site ($n,\downarrow$) lies  between two spin-up sites,
($n,\uparrow$) and ($n+1,\uparrow$).
This model 
 has therefore  the geometry of a frustrated zig-zag chain
as depicted in the lower half of Fig.\ \ref{hsplit}.

One should note that this system has a total of $2N$ sites, $N$
sites for each fermion species.
[Although it would be useful to adopt the notion
of a system with two coupled (spinless-fermion) 
chains, we will keep here the notation of a 
system of spinfull fermions.]

\begin{figure}
\epsfig{file=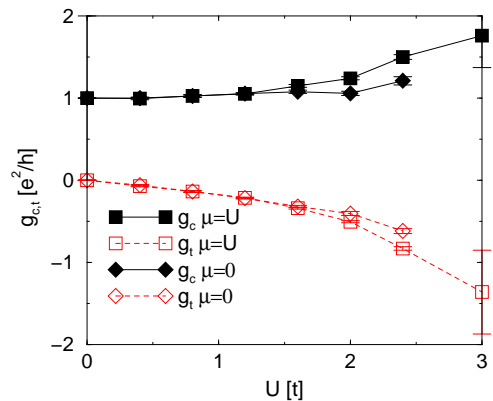,width=0.35\textwidth,angle=0}
\caption{Cis- and Transconductance (filled/empty symbols) of 
the zig-zag chain for two magnetic fields (120 sites per fermion
 species, $T=0.02t/k_B$, $U^\prime =U$ in Eq. (\ref{hamillad}), OBC's,
 $2\cdot 10^5$ MC sweeps.) }
\label{huls}\end{figure}

\begin{figure}
\epsfig{file=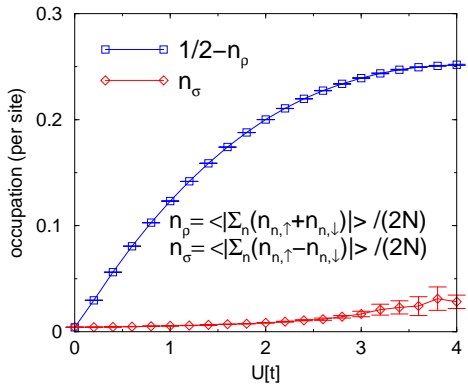,width=0.35\textwidth,angle=0}
\caption{Occupation per site  ($n_\rho$)
and difference between the occupations of the two fermion 
species ($n_\sigma$) 
for  the zig-zag chain.
 (100 sites per fermion species,
 PBC's, $\mu=0$, $10^4$ MC sweeps, $T= 0.1t$.) }
\label{stgmUzz}\end{figure}

%This model can be represented if we put 
% $U^\prime=U$ in Eq.\ (\ref{hamillad}). 
  The results for the spin drag in this model
 are shown in Fig.\ \ref{huls}.
 We discuss again two chemical potentials: one is $\mu=U$ implying half
filling,
 the other is again $\mu=0$.
% corresponds to the chemical potential shift---away from
% half filling---induced by the Coulomb interaction. 
In the latter case
the (mean) occupation number per site $n_\rho$
 is different from one half (the occupation at half filling) and depends on $U$.
It is shown in Fig.\ \ref{stgmUzz}. 

One sees in Fig.\ \ref{huls}
 that $|g_{c,t}|$ grow with the strength of the interaction.
This may be explained as follows: First, the Coulomb interaction mediates an
{\it attractive} nearest-neighbor interaction for fermions with equal
 spin orientation (this is a consequence of the frustration).
Therefore, in a simple approximation the only effect of the Coulomb 
interaction is to renormalize
the Luttinger-liquid parameters  for 
the two spin sectors
$K^{\uparrow,\downarrow}$.
Since the Luttinger parameter for a spinless-fermion chain increases with the 
strength of the attraction,\cite{GogNerTsve}
we expect that $K^{\uparrow,\downarrow}$ increases as $U$ increases.
Since $K^{\uparrow,\downarrow}$ gives  the conductance of 
one spin sector\cite{ApelRice} (which is essentially the Cisconductance)
 we have that $g_c$ increases with $U$.

One may also infer  from the figure that the dependence 
of $g_{c,t}$ on a chemical-potential shift is
weak. Within error bars $g_c$ decreases only slightly upon 
 shifting $\mu$  away from half filling.

\begin{figure}
\epsfig{file=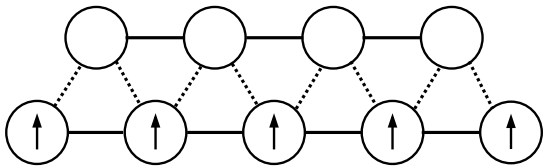,width=0.4\textwidth,angle=0}
\caption{One of two possible ground state configurations
for the zig-zag Hamiltonian
when $U>>t$. The sites in the lower row are occupied by spin-up
fermions, the  sites in the upper row are empty.
 In the other ground state, the lower row
is empty, and the upper completely filled with (spin-down) fermions.
}
\label{anfecofig}\end{figure}

One should note that
in the limit $U=\infty$ the ground state is a
spin-polarized
configuration (see Fig.\ \ref{anfecofig}).
  For $\mu=U$ this means that all conductances are 
zero in the large-$U$ limit, for one spin sector is
 empty and the other, completely filled. In contrast to this,
for $\mu= 0$ one of the two spin sectors may remain conducting.
The crossover to the ordered state occurs at values of 
 $U$ larger than $3t$ which may be seen by 
simulating and comparing the occupation number for different states.
For $\mu=0$
the difference in occupation (between the two spin sectors)
$n_\sigma=\langle|\sum_n \left(n_{n,\uparrow}-n_{n,\downarrow}\right)|\rangle/(2N)$ is shown in  
Fig.\ \ref{stgmUzz}; for $\mu=U$ it is zero within error bars as long as
$U\leq 3t$.
We conclude  that for the values of $U$ considered in Fig.\ \ref{huls}
the two spin sectors have approximately the same filling.
%%%%%%%%%%%%%%%%%%%%%%%%%%%%%%%%%%%%%%%%%%%%%%%%%%%%%%%%%%%%
\section{Conclusion}
In this paper we discussed the spin drag for the Hubbard model at
zero temperature.
We found that the Transconductance is negative---at half filling
the Umklapp even enforces $g_c=-g_t$.
In that respect our situation is different from  two 
coupled Tomonaga-Luttinger models
as considered in Refs.\ \onlinecite{Flens,KomnikEgger,NazarAver}
which do not incorporate Umklapp.
The  ``absolute-drag'' result of the form $g_c=g_t$
(e.g., from Ref.\ \onlinecite{NazarAver}) can only be recovered by
introducing a spin-polarized interaction (see appendix).

If we assume that a given potential is in general not spin dependent,
the only relevant quantity is the full conductance
$g=2(g_c+g_t)$, which is only nonzero away from half filling.
Here both spin orientations contribute equally to the current.
However, the situation changes when we add a magnetic impurity. 
Even if the driving potential is  still spin independent,
the resulting current will be (partially) spin polarized, if we are away
from half filling.
In the limit of a large impurity potential the current will be fully
spin polarized.

%Further, we 
%find that away from half filling a magnetic impurity closes the wire
%for one fermion species (up-spins/down-spins).
%%%%%%%%%%%%%%%%%%%%%%%%%%%%%%%%%%%%%%%%%%%%%%%%%%%%%%%
\begin{appendix}
\section{Spin-dependent interaction---broken spin-rotational invariance}
In the Hubbard model Cis- and Transconductance have opposite sign, in
sharp contrast to the bosonization results (for coupled spinless-fermion
chains), where Cis- and
Transcoductance
are both positive. The discrepancy may come from the different
modeling of the interaction. In the bosonization approaches each chain 
is given by
an interacting system (i.e, the Luttinger-liquid parameter $K$ may
differ from one), in the Hubbard model each spin sector
alone is represented by a noninteracting fermion system.
We will show in this appendix that a spin-polarized interaction leads
to a positive Transconductance as found in the bosonization approaches.
To this end we will now discuss the following variant of our Hamiltonian:
\begin{equation}\label{hubbjz}H=H_{\rm Hubb}+\sum_{n,\sigma}J_{z}(n_{n,\sigma}-1/2)(n_{n+1,\sigma}-1/2).
\end{equation}
Here the new $J_z$ term breaks the spin-rotational invariance.
Hence $K_\sigma$ may now be different from one.

%We use the transformation to a fermionic system used in Sec.\ \ref{sec_hubb}.
%($S_{2n}^z+1/2\to n_{n,\uparrow}$ and 
%$S_{2n+1}^z+1/2\to n_{n,\downarrow}$).
First we consider the {\it large-$U$ limit at half filling.}
The $U$ term acts then  as an effective projection to the configurations
with exactly (because of half filling) one fermion per site, i.e., $n_{n,\uparrow}=1-n_{n,\downarrow}$.
We now set-up an effective (second-order perturbation theory) Hamiltonian.
From the kinetic-energy term we get again a Heisenberg model with exchange
parameter $4t^2/U$.
The $J_z$ term of the Hamiltonian does not change the
configuration (in the occupation-number basis)
and gives therefore a direct energy contribution 
$2J_z\sum_nT_n^zT_{n+1}^z$ to the effective Hamiltonian where
$T_n^z=(n_{n,\uparrow}-n_{n,\downarrow})/2$
denotes the spin of the
 fermion on site $n$.
The full effective Hamiltonian reads 
\begin{eqnarray*}
H_{\rm eff}&=&\sum_n(4t^2/U)(T_n^+T_{n+1}^-+T_{n+1}^+T_{n}^-)/2\\
&+&\sum_n
(2J_z+4t^2/U)T_n^zT_{n+1}^z,\end{eqnarray*}
and is an $xxz$ chain. If the anisotropy is larger than
the
hopping amplitude, i.e., if $J_z>0$, this model is gapped (implying both
 a charge and a spin gap in the original model).
 We therefore expect that Cis- and
Transconductance go to zero, if we increase $U$ and keep a finite $J_z $.

Now we consider  a {\it zero chemical potential}  
 $\mu=0$. We expect that this chemical potential shift away from half
 filling closes the
 charge gap, but leaves the spin gap more or less unaffected.
We consider again the {\it large-$U$ limit}.
In any configuration the $J_z$ term of the
Hamiltonian
gives the following contribution for two neighboring sites
\begin{itemize}
\item[$-J_z/2$] if the two sites are occupied with anti-parallel spins,
\item[$J_z/2$] if the two sites are occupied with parallel spins or are both
empty,  
\item[$0$] if one site is occupied and the other, not.
\end{itemize}
We assume that there is a spin gap and
that the (degenerate) ground state  configurations are those 
for  which 
the spins of the particles are ordered antiferromagnetically.
If  only these configurations are allowed, the $J_z$ term
can be represented as a one-site potential with a contribution
$\pm J_z/2$  for empty/occupied sites. (One obtains the same energy
contributions as from the $J_z$ term of the original Hamiltonian,
if one keeps in mind that each site appears in precisely two pairs
of neighboring sites.) 
We can set-up the following effective Hamiltonian (this is just the
restriction of the original Hamiltonian to the assumed ground state
configurations, i.e.,
zeroth order in $U$)
$$\sum_n[(R_n^+R_{n+1}^-+R_{n+1}^+R_{n}^-)/2+J_zR_n^z],$$
which is an $xx$ chain in magnetic field,
where the ``spin'' operator $R_n^z$ (this time denoted by $R$ to avoid
confusion with previous spin operators) is defined by
$R_n^z=-1/2$ if there is a particle on site $n$
and $R_n^z=+1/2$ if site $n$ is empty. 
Since the effective Hamiltonian describes the charge part of the
Hamiltonian
 the full conductance $2(g_c+g_t)$ for the original model should
coincide with the conductance of the new Hamiltonian which is $e^2/h$
as the system is noninteracting.\cite{GogNerTsve,ApelRice}
Since we have $g_c=g_t$ by the assumption of a
spin gap and Eq.\ (\ref{Krs}), the relation $g=2(g_c+g_t)$
yields $g_c=0.25\, e^2/h=g_t$.

In principle the model Eq.\ (\ref{hubbjz})
 can also be analyzed with the  Monte Carlo
method developed in this paper,
but we found that the simulation for this case is problematic:
  We measured  large
autocorrelation times for finite $J_z$ and $\mu$ 
(e.g., for the computation of the compressibility).
We therefore must restrict ourselves to $J_z\leq 0.8t$.

\begin{figure}
\epsfig{file=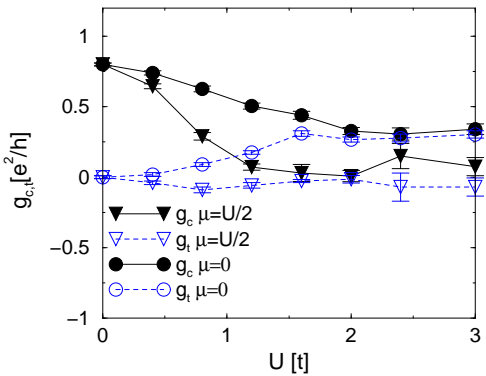,width=0.35\textwidth,angle=0}
\caption{The same as Fig.\ \ref{hullz1a}, but for $J_z=0.8t$ (here $N=140$).
The simulations away from half filling suffer from large autocorrelation times.
}
\label{hullz1b}\end{figure}

For $J_z=0.8t$  we present results for the Cis- and Transconductance
 in Fig.\ \ref{hullz1b}. 
In the large-$U$ limit we find good agreement with our prediction
that $g_c=g_t=0.25\, e^2/h$
which gives  credit to the simulation data despite the large autocorrelation times.

Here we want to stress once again
the remarkable fact that the sign of the
Transconductance (the direction of the induced current) changes when we
switch the magnetic field and the spin-polarized interaction on.
(The Transconductance is for all $U$ negative in Fig.\ \ref{hullz1a}
whereas in the present situation we expect $g_t=g_c=K_\rho/2>0$ for $T=0,U=\infty$.)

\begin{figure}[b]
\epsfig{file=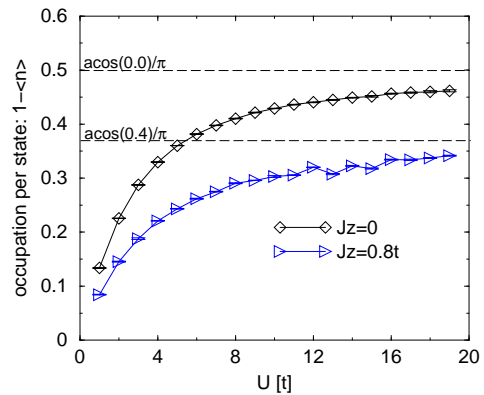,width=0.35\textwidth,angle=0}
\caption{Occupation per state (away from half filling) for  the Hamiltonian Eq.\ (\ref{hubbjz})
 $H(\mu=0)$  for different $J_z$.
The predicted high $U$ values are given as dashed lines. (500 sites,
 PBC's, $10^5$ MC sweeps, $T= 0.1t$.) }
\label{stgmU}\end{figure}

{\it Occupation in the ground state}---Since we identified the
 ground state of the Hamiltonian
Eq.\ (\ref{hubbjz}) 
 $H(\mu=0,U\to \infty)$
with the ground state of the $xx$ chain in magnetic field, we can 
calculate 
the occupation per state of this
Hamiltonian in the large $U$-limit, the result being:
$$\sum_{n,\sigma}(n_{n,\sigma})/N=1-{\arccos(J_z/[2t])}/\pi.$$
This prediction may be tested against a Monte Carlo simulation.
 We find good agreement 
(see Fig.\ \ref{stgmU}).
\end{appendix}

%%%%%%%%%%%%%%%%%%%%%%%%%%%%%%%%%%%%%%%%%%%

\end{document}